\documentclass[
aip,
apl,
graphicx,
reprint
]{revtex4-1}

\usepackage[T1]{fontenc}
\usepackage{graphicx}
\usepackage{hyperref}

\draft 

\begin{document}


\title{Impact of multi-dot configurations on electron temperature and double quantum dot latching in gate-controlled qubit devices}



\author{Jakob Walsh}
\affiliation{Fakultät für Physik, Universität Regensburg, 93053 Regensburg, Germany}
\author{Miriam Schweinböck}
\affiliation{Fakultät für Physik, Universität Regensburg, 93053 Regensburg, Germany}
\author{Jhih-Sian Tu}
\affiliation{Helmholtz Nano Facility (HNF), Forschungszentrum Jülich GmbH, 52428 Jülich, Germany}
\author{Yuji Yamamoto}
\affiliation{IHP - Leibniz Institute for High Performance Microelectronics, 15236 Frankfurt (Oder), Germany}
\author{Felix Reichmann}
\affiliation{IHP - Leibniz Institute for High Performance Microelectronics, 15236 Frankfurt (Oder), Germany}
\author{Lars R. Schreiber}
\affiliation{JARA-FIT Institute for Quantum Information, Forschungszentrum Jülich GmbH and RWTH Aachen University, 52062 Aachen, Germany}
\affiliation{ARQUE Systems GmbH, 52074 Aachen, Germany}
\author{Dominique Bougeard}
\email[]{dominique.bougeard@ur.de}
\affiliation{Fakultät für Physik, Universität Regensburg, 93053 Regensburg, Germany}


\date{\today}

\begin{abstract}
Many currently envisaged scalable architectures for Si/SiGe spin qubit devices utilize gate-controlled multi-dot configurations characterized by their peculiar Coulomb blockade physics. One example is the concurrent use of a single electron transistor (SET) as a charge sensor and as a reservoir for adjacent qubit quantum dots (QD). We find that the SET energy spectrum impacts pulsed gate spectroscopy of the QD when the SET simultaneously acts as the reservoir and the sensor. Then, using the qubit QD as a probe, we quantify how strongly the electron temperature of the SET depends on the applied bias. It surpasses a reference experiment by 100\,mK even at 50\,µV bias. Our results provide a strong incentive for choosing spin qubit initialization schemes which do not rely on coupling to the reservoir in such devices. Switching to a double quantum dot connected to a single reservoir, a frequent configuration for qubit readout, we finally demonstrate the tunability of the characteristic latching behavior by systematically mapping it in the few-electron stability diagram.
\end{abstract}

\pacs{}

\maketitle 

With two-qubit gate fidelities in Si/SiGe having reached the error correction threshold, \cite{Noiri.2022, Xue.2022, Wu.} scaling the amount of spin qubits and designing quantum processor architectures for universal quantum computing is receiving strong attention. \cite{Tadokoro.2021, Philips.2022, Ha.2022, Kunne.2024, George.2025} In this context, two approaches using different multi-dot configurations during operation have emerged: 2D qubit arrays and architectures leveraging qubit shuttling. \cite{Kunne.2024, Unseld.2025, Undseth.2026, Tosato.2026} While the Coulomb blockade physics of such configurations goes beyond textbook descriptions, it has practical consequences for the qubit initialization and readout. One prominent example are integrated charge sensors for baseband detection. To reduce the number of gates an integrated single electron transistor~(SET) will cover two functionalities for the adjacent quantum dots (QD): In addition to charge sensing, it simultaneously also plays the role of a charge carrier reservoir. \cite{Struck.2024, Tosato.2026, FernandezdeFuentes.2026, Wu.} A double quantum dot (DQD) coupled to only one reservoir, as frequently used in compact settings of Pauli spin blockade (PSB) readout, is a second example for which a peculiar latching behavior has been reported. \cite{Yang.2014, HarveyCollard.2018}

In this letter, we first discuss Coulomb blockade signatures that emerge when an SET is used in a dual \textit{reservoir-sensor} role in a 1D array. In particular, we quantify the impact on the electron temperature as a measure of the single electron transition broadening of an adjacent QD when using the SET as a reservoir. We then demonstrate how our gate layout provides a broad tunability of the latching behavior in a DQD only connected to a single reservoir.

All measurements were conducted in a dilution refrigerator with a base temperature below 10 mK. The cryostat's DC lines are thermally anchored at every temperature stage (50\,K, 4\,K, 1\,K, 100\,mK and 10\,mK) via gold-plated copper bobbins and are equipped with two-stage cryo filters at the mixing plate: A radio frequency low-pass LC filter bank and a low frequency low-pass RC filter bank attenuate above 225\,MHz and 65\,kHz, respectively. An I/V converter is used for base-band readout of the SET current.

\begin{figure}
    \centering
    \includegraphics[width=0.47\textwidth]{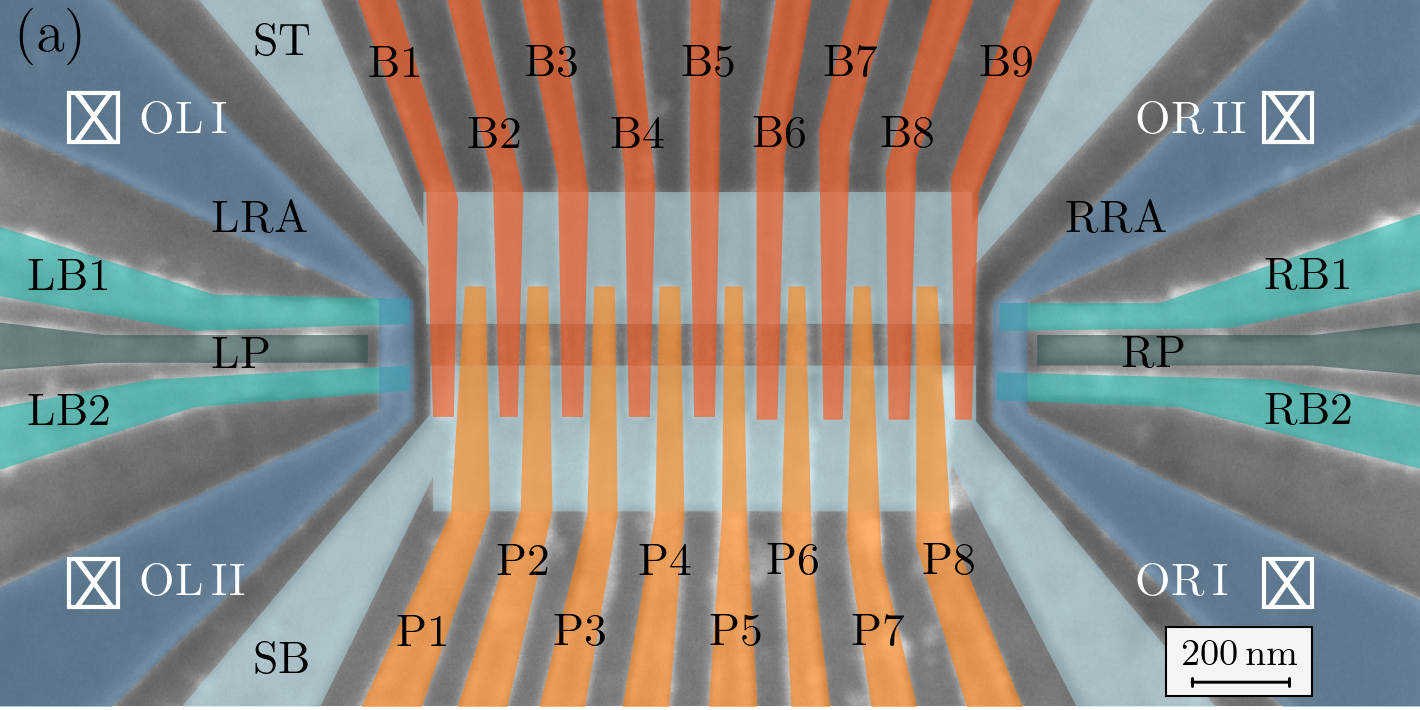}
    \vfill
    \includegraphics[width=0.23\textwidth]{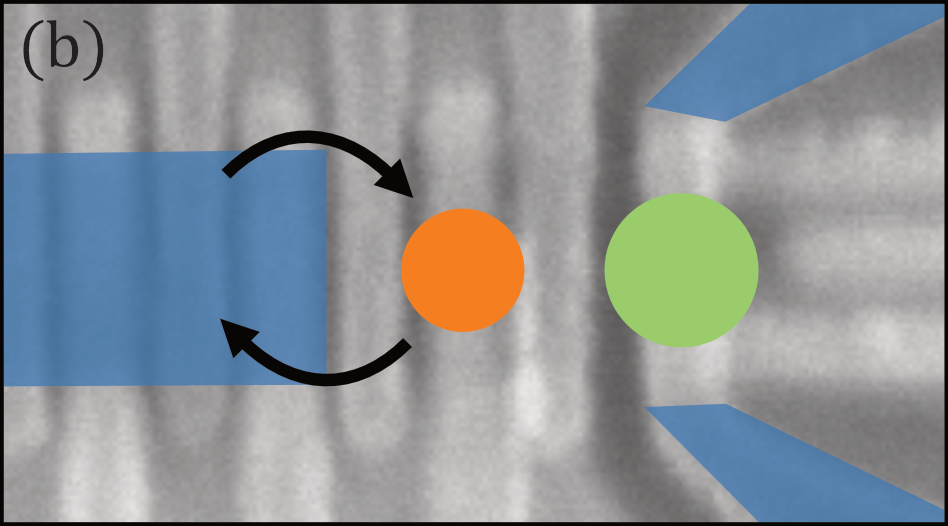}
    \includegraphics[width=0.23\textwidth]{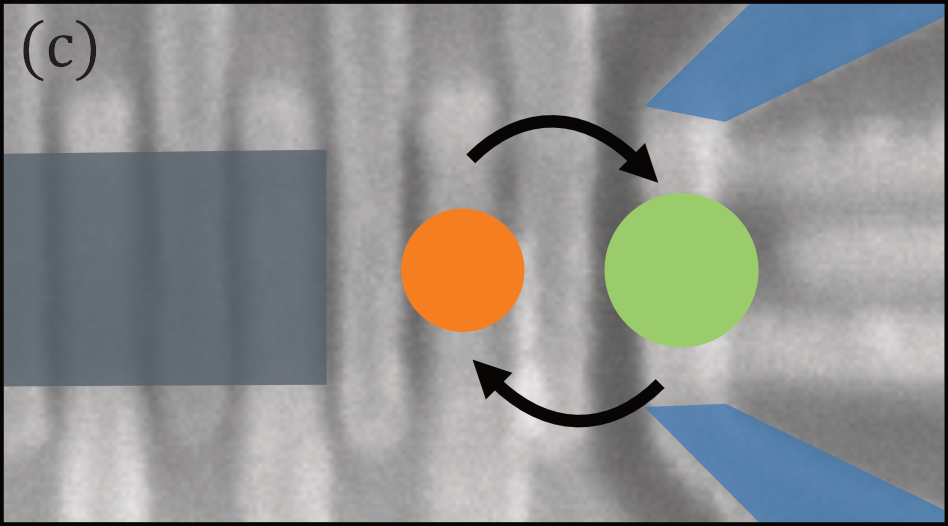}
    \caption{\label{fig:layout} (a) SEM of device A with the individual gates and reservoirs marked in color. Each color indicates similar functionality. (b),(c) SEM with the positions of each reservoir (colored in blue), the SET island (green) and the QD (orange) for the two measured configurations. The coupling to the \textit{extended reservoir} (b) and to the \textit{reservoir-sensor} (c) which differentiates the two measured configurations is indicated via arrows.}
\end{figure}

Two devices, A and B with identical layouts, were used for the measurements presented in this work. Figure~\ref{fig:layout} shows a false color scanning electron micrograph (SEM) of device A. The heterostructure grown by chemical vapor deposition (CVD) consists of an 11\,nm thick Si quantum well between $Si_{0.67}Ge_{0.33}$ barriers. The top barrier being 35\,nm thick. The Ti/Pd gates are distributed across three layers, each separated by nominally 10\,nm of atomic layer deposition (ALD) AlOx. This type of layout has been employed for shuttling experiments. \cite{Seidler.2022, Volmer.2024, Struck.2024, Xue.2024, Beer.} This is reflected in a channel delimited by screening gates in light gray and modulated by individually controllable claviature gates in orange and red. At either end of the channel two barriers and a plunger gate enable the formation of an SET. The SET and its reservoirs connected to two ohmic contacts are accumulated by a common top gate. Device B features the same gate layout. The only difference is that the QW thickness is 13\,nm. 

The layout implies that one or more QDs can be deterministically formed within the channel. The SET at each end of the channel plays two roles: They simultaneously serve as a charge sensor for the QDs in the channel and also as reservoirs for the QDs. \cite{Klos.2018} This double \textit{reservoir-sensor} role differs from typical textbook situations, where the QDs are initialized via dedicated \textit{extended reservoirs} which are spatially separated from baseband charge sensors. To investigate the double role of a \textit{reservoir-sensor} SET, we form a QD below P8 with barriers B8 and B9 of device A, the gates being indicated in Figure~\ref{fig:layout}\,(a). RB1 and RB2 form the barriers of the SET. Furthermore, we use the remaining claviature gates to accumulate an \textit{extended reservoir} on the left side of the QD. This enables the two different loading modalities that are sketched in Figures~\ref{fig:layout}\,(b) and \ref{fig:layout}\,(c), respectively: Selectively couple to the QD either from the \textit{extended reservoir} on the left or from the \textit{reservoir-sensor} SET in its double role on the right side. Thus, we have access to two operation modalities: Either, operating reservoir and SET separately or using the \textit{reservoir-sensor} mode.

\begin{figure*}
    \centering
    \includegraphics[width=0.49\textwidth]{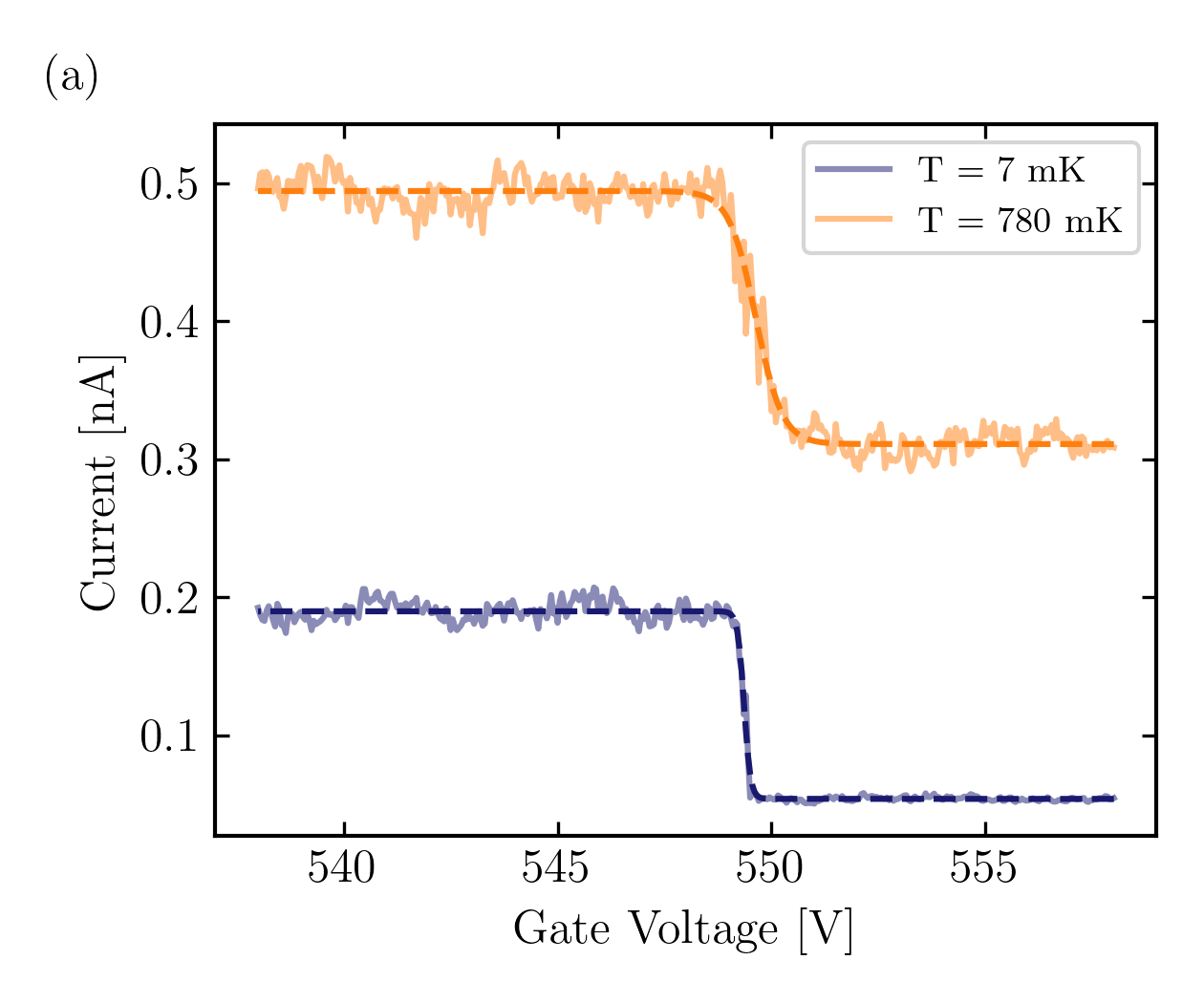}
    \includegraphics[width=0.49\textwidth]{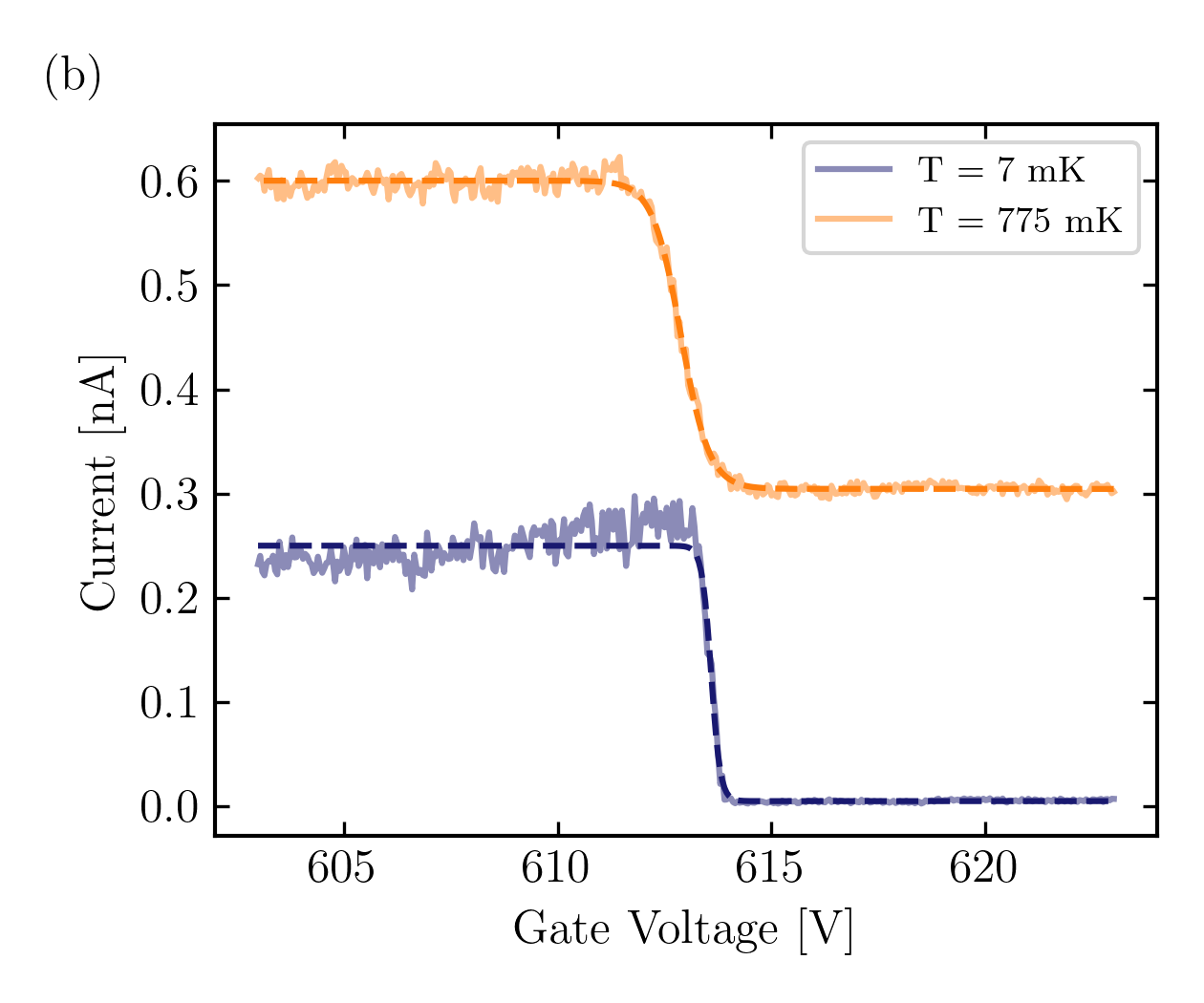}
    \includegraphics[width=0.49\textwidth]{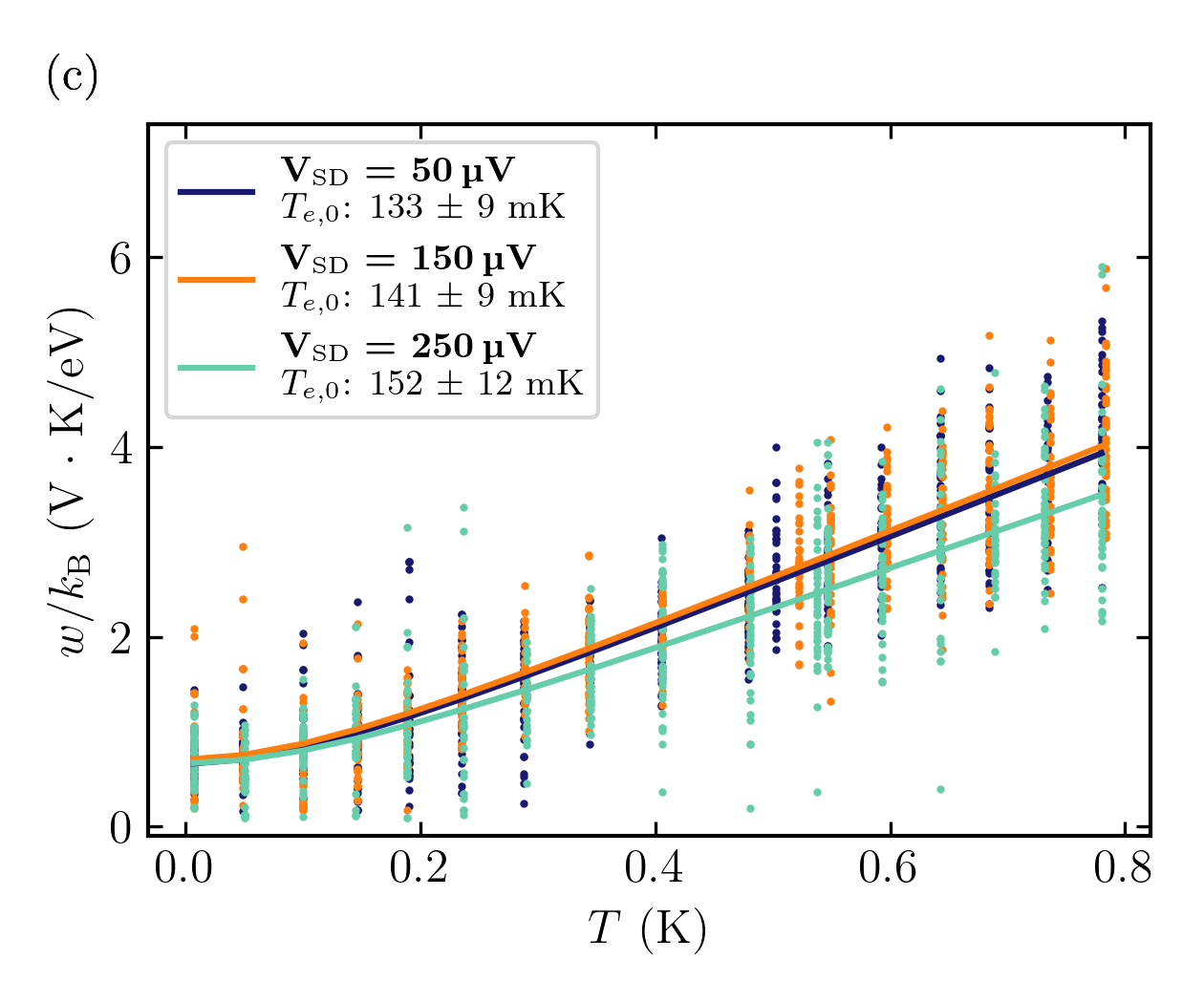}
    \includegraphics[width=0.49\textwidth]{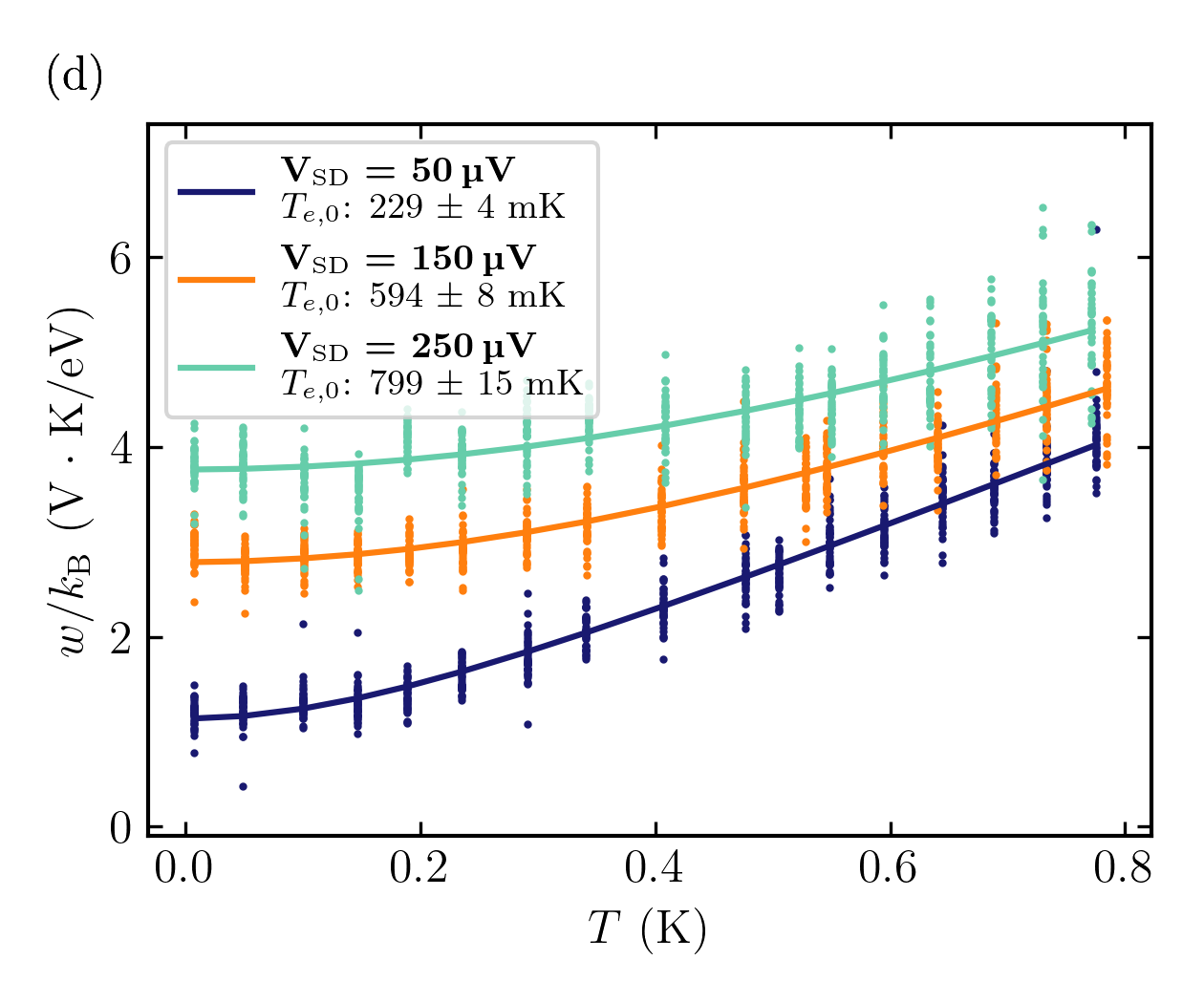}
    \caption{\label{fig:elec_temp} Individual $0-1$ QD transitions with Fermi fit according to Eq.~\ref{eq:fermi} for the loading from (a) the \textit{extended reservoir} and from (b) the \textit{reservoir-sensor} SET for 50\,µV SET bias. For clarity, the traces are shifted vertically. Fermi fit broadening of the $0-1$ QD transition fit as a function of the mixing plate temperature T and for three different (50\,µV, 150\,µV and 250\,µV) SET bias voltages. The data is fitted (Eq.~\ref{eq:temp}) to extract the electron temperature and lever arm for the loading from (c) the \textit{extended reservoir} and from (d) the \textit{reservoir-sensor} SET. The extracted lever arms are 201\,meV\!/V, 198\,meV\!/V and 227\,meV\!/V for the \textit{extended reservoir} and 201\,meV\!/V, 213\,meV\!/V and 212\,meV\!/V for the \textit{reservoir-sensor} SET configuration.}
\end{figure*}

We start by determining the electron temperature of both types of reservoirs, which is a metric for the Fermi-Dirac distribution in the respective reservoir. Experimentally, we track the $0-1$ transition of the tunnel-coupled QD to probe this energy distribution. Since the tunnel coupling of the QD is below 50\,kHz, lifetime broadening effects are on the order of magnitude of µK and are therefore negligible. \cite{Gasparinetti.2012} Figure~\ref{fig:elec_temp}\,(a) and (b) show exemplary virtualized transition traces fitted with a Fermi function \cite{Gasparinetti.2012, Beenakker.1991}
\begin{equation}
I_{SET} = \frac{a}{e^{\left(V - V_0\right) / w} + 1} + b,
\label{eq:fermi}
\end{equation}
for specific mixing plate temperatures. Here, $V$ is the gate voltage of P8, $V_0$ is the charge transition point and $a$ and $b$ account for the current background. The width of the transition is denoted as $w = \frac{k_B \times T_{e0}}{\alpha}$. $\alpha$ is the lever arm of P8 and $T_{e0}$ is the electron temperature of the respective reservoir. The width of the Fermi function as a function of the mixing plate temperature $T$ is phenomenologically described by \cite{Prus.2002, Simmons.2011}
\begin{equation}
w(T) = \frac{k_B}{\alpha} \sqrt{T^2 + T^2_{e0}},
\label{eq:temp}
\end{equation}
and thus allows for the extraction of the lever arm $\alpha$ and the electron temperature $T_{e0}$ when experimentally varying the mixing plate temperature.

First, we extract the electron temperature of the \textit{extended reservoir}, with the reservoir potential set to 0\,V (Figure~\ref{fig:layout}\,(b)). Barrier B9 is tuned to an opaque setting such that QD and SET are purely capacitively coupled. Figure~\ref{fig:elec_temp}\,(c) shows the width of the Fermi fits for mixing plate temperatures up to 800\,mK for three different SET bias voltages. The extracted electron temperature increases from 133\,mK for 50\,µV bias voltage to 152\,mK for 250\,µV bias voltage. Hence, the electron temperature increases only weakly with the SET bias voltage and the SET current. At these low temperatures, the electron temperature of the \textit{extended reservoir} is mostly decoupled from the crystal temperature. \cite{Wellstood.1994} The measured electron temperatures correspond well with reports on devices that are also initialized via dedicated \textit{extended reservoirs}. \cite{Lai.2011, Simmons.2011, Maune.2012, Dodson.2020, Hollmann.2020, Neyens.2024, Marcks.2025}

We now change the barrier tuning of the QD such that the SET is used simultaneously as a reservoir and as the charge sensor, while ensuring that the tunnel coupling is the same as for the \textit{extended reservoir} case. Also, the SET barrier voltages are adapted to ensure a similar SET current as before. The \textit{extended reservoir} is no longer coupled to the QD (Figure~\ref{fig:layout}\,(c)). Therefore, all electrons that are now loaded into the QD originate from the \textit{reservoir-sensor} SET. As a consequence, we can now quantify the electron temperature of the SET island, which is operated in a many-electron regime where spatial confinement is negligible. The results of the electron temperature measurements for the same bias voltages as before are presented in Figure~\ref{fig:elec_temp}\,(d). For the lowest bias voltage of 50\,µV, the electron temperature is 229\,mK. For 250\,µV bias it increases to 799\,mK.
Hence, the electron temperature of a biased SET (used in the \textit{reservoir-sensor} operation mode) is approximately 100\,mK higher than the one of an unbiased reservoir (used in the \textit {extended reservoir} operation mode). The experiment also quantifies the rapid increase of electron temperature with SET bias.

The usage of the SET as a charge sensor requires appropriate Coulomb blockade peaks, which emerge only when a finite bias voltage is applied between the source and drain reservoirs of the tunnel-coupled SET. This bias, and possibly the tunnel rates to source and drain, are responsible for a broadening of the Fermi-Dirac distribution in the \textit{reservoir-sensor} SET, leading to the observed higher electron temperature. Given that the single electron QD probes this broadening of the Fermi-Dirac distribution, our experiment quantifies the bias-dependence of the resulting QD transition broadening in this operation mode. The current across the SET was kept the same for both electron temperature measurements and the QD position is only slightly shifted due to the different barrier tuning. When the QD is loaded from the \textit{extended reservoir}, which has zero voltage applied, electrons closer to equilibrium tunnel into the QD. The QD then probes the lower electron temperature of this reservoir. The SET current produces Joule heating, which may affect its environment, including the distant \textit{extended reservoir}. We attribute the small SET bias dependence of the electron temperature in the \textit{extended reservoir} to Joule heating originating from the SET. \cite{Korotkov.1994} The comparison of the electron temperatures shows that when the \textit{reservoir-sensor} mode is used in a qubit device, an optimization of the SET bias (and/or the SET current) is required, to minimize the transition broadening probed by the QD while maintaining charge sensitivity.

\begin{figure*}
    \centering
    \includegraphics[width=0.49\textwidth]{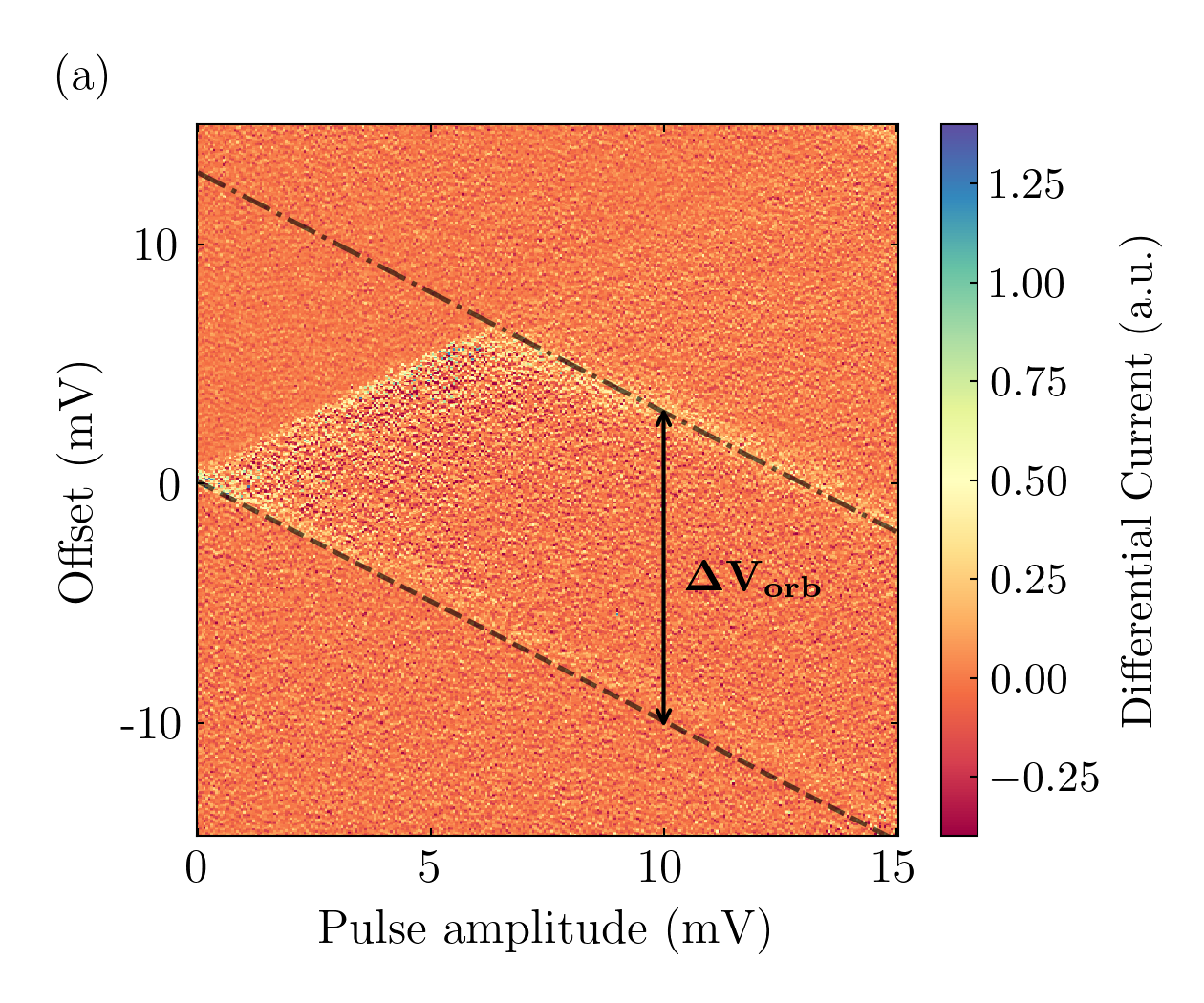}
    \includegraphics[width=0.49\textwidth]{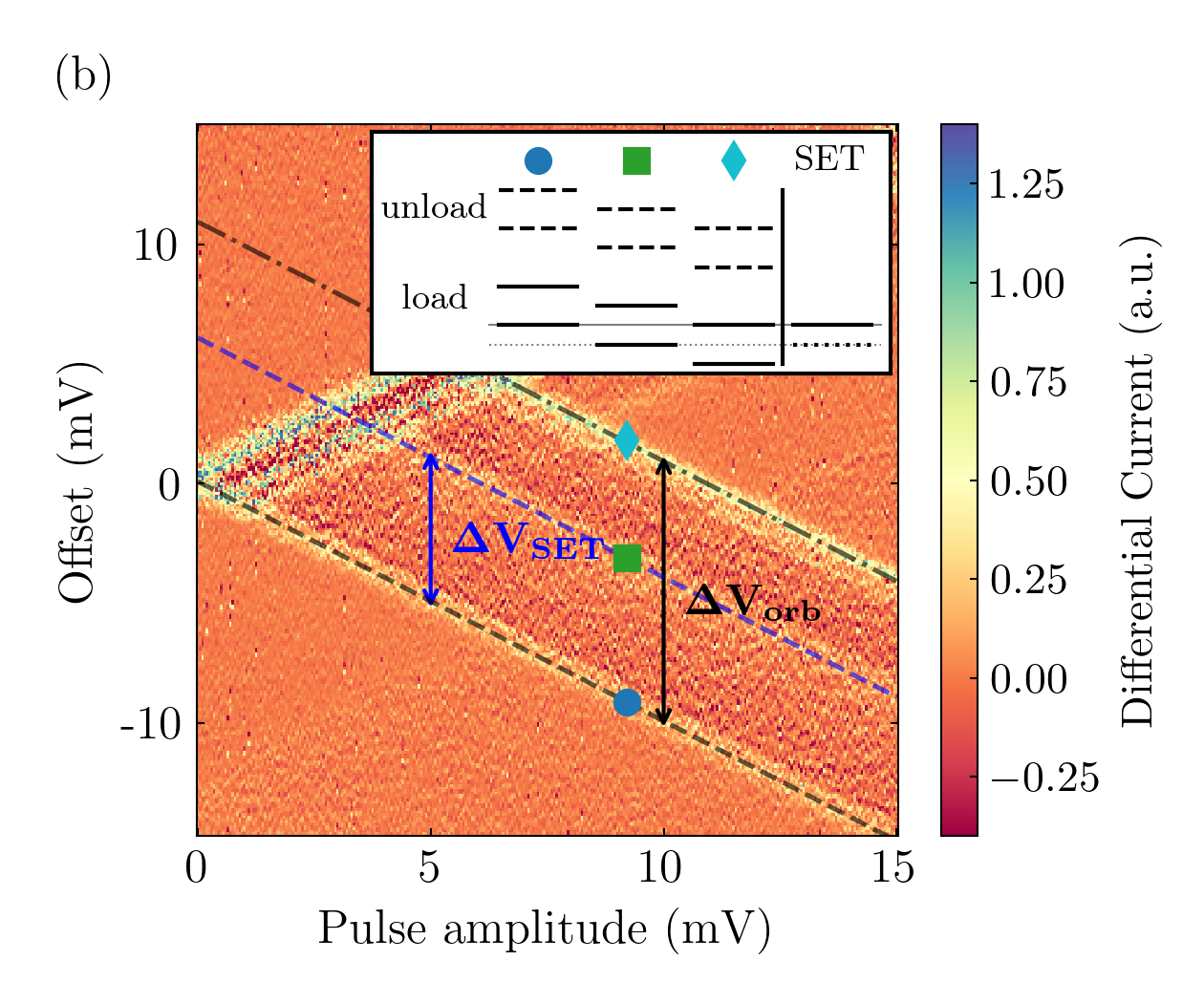}
    \caption{\label{fig:pgs} Single electron pulsed gate spectra of the QD for the two loading modalities from (a) the \textit{extended reservoir} and (b) the \textit{reservoir-sensor} SET. The inset shows the energies of the ground and the first orbital state during the loading (solid line) and unloading (dashed line) parts of the pulse for three positions marked in the main panel. The alignment with the $N$ (solid line) and $N-1$ (dotted line) electron in the SET is indicated via gray lines.}
\end{figure*}

In a next step, we study the impact of the SET operation mode on the single-electron energy spectrum of the QD obtained from pulsed gate spectroscopy (PGS). \cite{Elzerman.2004} The lever arms to convert the voltages into energies are extracted from the mixing plate temperature-dependent transition broadenings in Figure~\ref{fig:elec_temp}. Figure~\ref{fig:pgs}\,(a) shows a PGS measurement of the $0-1$ transition of the QD in the \textit{extended reservoir} configuration. An energy state 2.9\,meV above the ground state is clearly visible. It matches the expected energy of the first orbital QD state for these devices. \cite{Volmer.2024} Notably, no state that could be attributed to the valley splitting is visible. Separate, magnetospectroscopy measurements (not shown) indicate a valley splitting below 50\,µeV and thus below the resolution of these PGS measurements. Switching to the \textit{reservoir-sensor} SET configuration and repeating the measurement in Figure~\ref{fig:pgs}\,(b), now reveals broader resonance lines due to the increased electron temperature. Two loading lines are visible and they correspond to energy states 2.3\,meV and 1.3\,meV above the ground state. We attribute the energy of 2.3\,meV again to the first orbital state of the QD. The small deviation in orbital energies between both reservoir configurations results from the difference in barrier voltages B8 and B9, slightly changing the QD confinement potential. The energy state at 1.3\,meV is too low to represent an orbital state of the QD. Notably, the energy of 1.3\,meV actually closely matches the SET charging energy of 1.4\,meV extracted from Coulomb diamonds.

We understand the occurrence of the energy state in the PGS measurement as the result of the alignment of the QD ground state with the $N-1$ state of the SET. This is sketched in the inset of Figure~\ref{fig:pgs}\,(b). The PGS method is based on a square wave pulse that repeatedly loads and unloads the QD with a frequency similar to that of the tunnel rate. When the tunnel rate increases as a result of the alignment of additional QD energy states with the reservoir, the average dot occupation increases as well. However, the tunnel rate is also enhanced when the number of initial states in the reservoir increases. For the \textit{reservoir-sensor} SET configuration this is the case when in addition to the $N$ SET electron also the $N-1$ SET electron is available for loading. Thus, we conclude that the energy state in the PGS measurement corresponds to the difference between the $N$ and the $N-1$ electron states of the SET i.e., the charging energy of the SET. The comparison of the two operation modes of the SET hence quite directly illustrates how the Coulomb blockade island nature of the reservoir from which the QD is loaded will manifest in a PGS experiment and affect the loading of the QD. Note that the origin of the additional unloading lines occurring in the \textit{reservoir-sensor} SET configuration in Figure~\ref{fig:pgs}\,(b) is yet unclear.

\begin{figure*}
    \centering
    \includegraphics[width=0.98\textwidth]{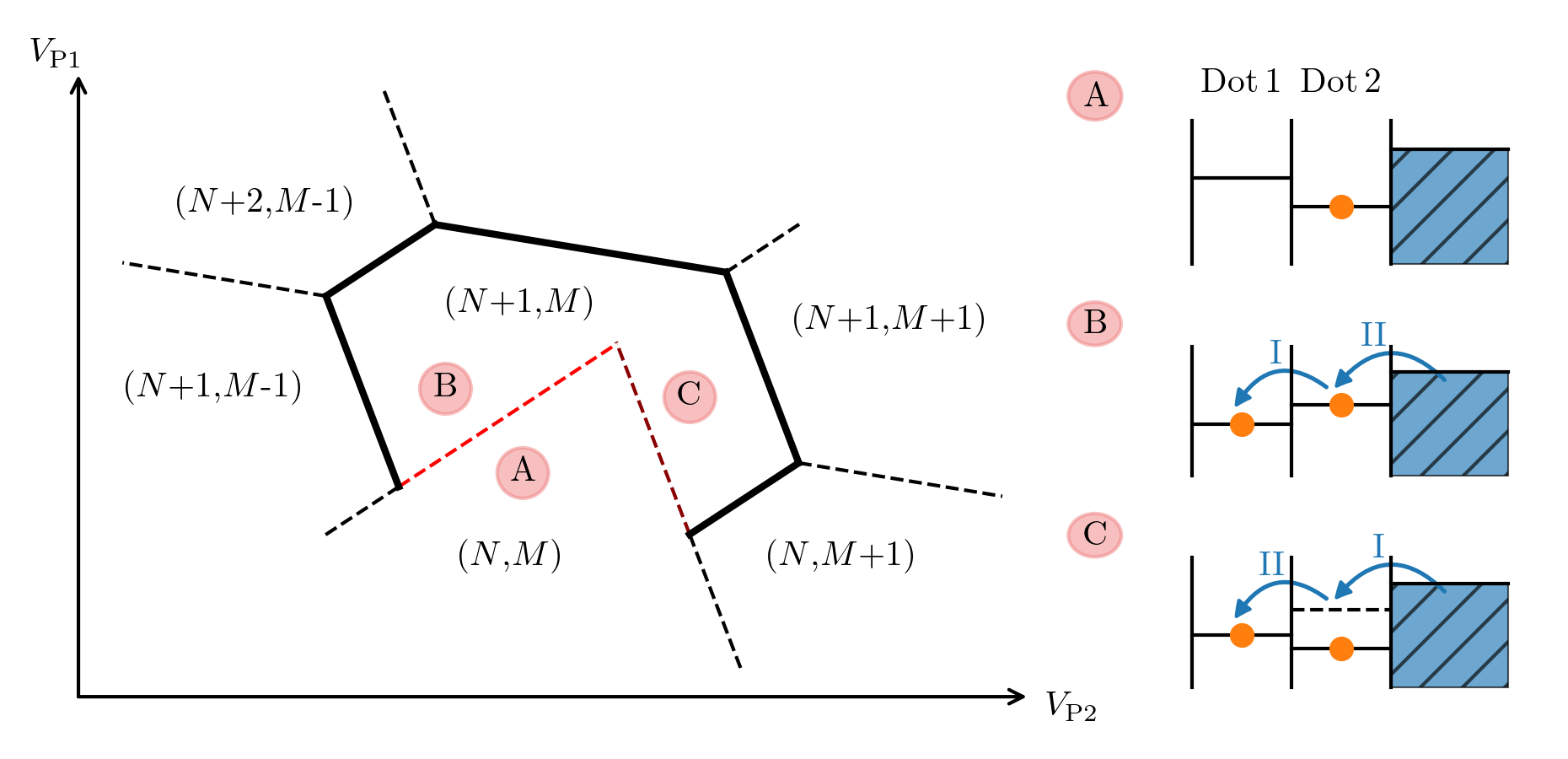}
    \caption{\label{fig:dqd_stab} Schematic unit cell of the DQD stability diagram with latching: The latched transitions for the loading behavior are marked in light and dark red. Selected potential schemes of the DQD and the reservoir are sketched on the right for the indicated areas A, B, C close to the latched transitions. When applicable, the electron tunneling sequence is indicated with arrows.}
\end{figure*}

Another frequent operation mode in scalable qubit architectures is to connect a DQD solely to the \textit{reservoir-sensor}. In shuttling devices this enables spin-to-charge conversion via PSB of a shuttled electron. \cite{Volmer.2024} Here, we discuss the DQD Coulomb physics resulting from the use of only one reservoir in device B. The DQD, consisting of QD1 and QD2, is formed on the left side of the device below P1 and P2, respectively. B1, B2 and B3 are acting as barriers. B1 is set sufficiently opaque so that there is no tunnel coupling to the SET. Therefore, only QD2 is directly coupled to the \textit{extended reservoir} via B3. In general, this leads to a latching signature in the DQD stability diagram first described by Yang et al. \cite{Yang.2014} which we shortly review in Figure~\ref{fig:dqd_stab}. In the schematic unit cell of the DQD stability diagram, instead of the horizontal loading line associated with the direct loading of QD1, the latching signature manifests as a triangular extension of the charge stability region \textbf{A}. This region is formed by the continuation of the diagonal interdot and the vertical QD2 transitions. Indeed, the triangular shape is a direct consequence of the missing left reservoir: As shown in situations \textbf{B} and \textbf{C} in Figure~\ref{fig:dqd_stab}, QD1 can only be loaded from the right reservoir via mediating energy states of QD2.

\begin{figure*}
    \centering
    \includegraphics[width=0.98\textwidth]{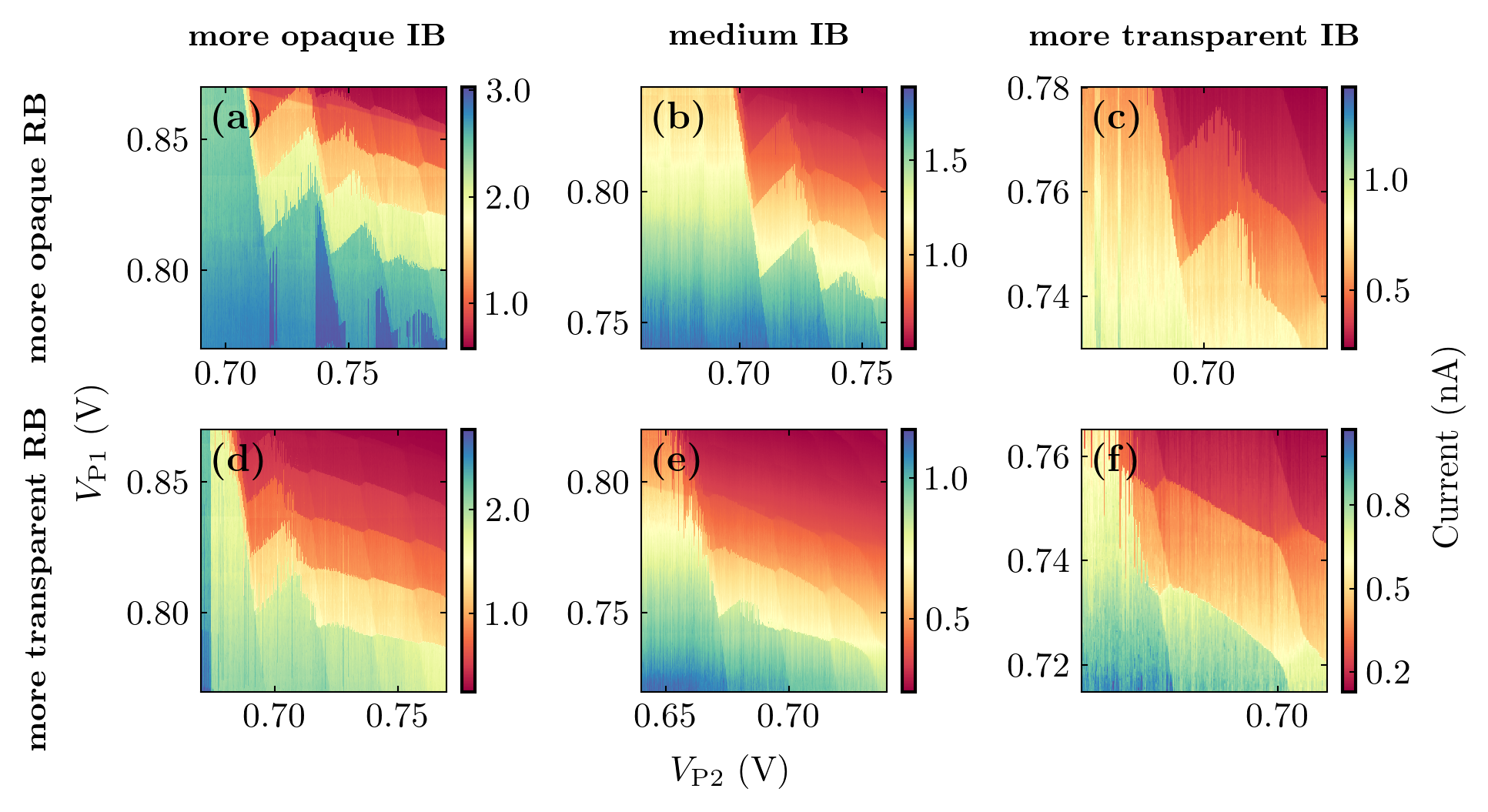}
    \caption{\label{fig:dqd_tuning} Experimental DQD stability diagrams exhibiting the latching behavior for combinations of gate-tuned (a) more opaque RB and IB, (b) more opaque RB and medium IB, (c) more opaque RB and more transparent IB, (d) more transparent RB and more opaque IB, (e) more transparent RB and medium IB and (f) more transparent RB and IB tuning.}
\end{figure*}

In the following, we show that our device layout permits a broad tuning of the latching of the DQD stability diagram. To this end, we tune the interdot barrier (IB) and the reservoir barrier (RB) while performing DQD stability measurements. IB corresponds to gate B2, while RB is tuned via P3 and B4 instead of B3 to avoid stronger cross-coupling effects. The results are presented in Figure~\ref{fig:dqd_tuning}. For fairly opaque IB and RB values in Figure~\ref{fig:dqd_tuning}\,(a), we find that the latching affects all (N,1) and also (N,2) transitions while it vanishes for higher electron occupations of QD2. For higher electron occupations of QD2 the cross-coupling of the plunger gate results in more transparent IB and RB. As a result QD1 is now coupled directly to the \textit{extended reservoir} on the right and loading of electrons can occur without the need of QD2 states. Note, that such direct coupling does not occur for higher electron occupations of QD1, since here the cross-coupling only affects IB, resulting in a much weaker increase of the direct tunnel coupling. The occurrence of the latching directly depends on the measurement speed: The latching effect will occur as long as the measurement speed is faster than the rate of direct coupling between QD1 and reservoir. Given the integration time per data point of 20\,ms in our experiment, latching occurs if the direct tunneling times remain longer than 20\,ms. Now, tuning solely IB gradually to more transparency in Figure~\ref{fig:dqd_tuning}\,(b),\,(c), leaving RB opaque, the extent of the latching may be reduced to only occur for the (N,1) transitions. Tuning RB to more transparency enables an even larger tuning range of the latching extent as seen in Figure~\ref{fig:dqd_tuning}\,(d) to (f): In Figure~\ref{fig:dqd_tuning}\,(f), the latching effect disappears completely for the most transparent IB and RB. The $0-1$ transition of QD1 is now directly coupled to the reservoir. Figure~\ref{fig:dqd_tuning}\,(e) shows an intermediate regime in which the latching mechanism is only visible for part of the $0-1$ transition of QD1. For lower voltage values of P2 the latched interdot transition dominates. However, when P2 is increased, the strength of the direct coupling of QD1 and the reservoir increases and the direct transition becomes available. First, this is only facilitated via the increased tunneling rate into the orbital state. \cite{Simmons.2011} It results in the typical fringe-like transition. For higher P2 the ground state tunnel rate then becomes sufficiently high and the transition stabilizes.

In conclusion, we have documented the relevant CB physics in two multi-dot situations occurring in scalable spin qubit architectures. In a configuration where a DQD is connected to only one reservoir in the device, we have revisited the characteristic latching behavior, by systematically exploring its parameter space as a function of the relevant tunneling couplings, using the broad tunability allowed by the gate layout.
In a second multi-dot situation, where a single electron qubit QD is adjacent to a dual role \textit{reservoir-sensor} SET, the resulting impact of the discretized energy spectrum of the SET on the loading behavior of the qubit QD can be quantified in PGS measurements.
Additionally, we show that the electron temperature probed by a qubit QD will be significantly impacted by the bias voltage of a \textit{reservoir-sensor} SET. It strongly increases with the bias of the adjacent SET. Our study highlights the importance of optimizing the bias for the \textit{reservoir-sensor} SET. This will be particularly true during the initial tuning of qubit QDs and for QD spectroscopy. Furthermore, it will impact implementations of high-fidelity readout schemes in device architectures where biased SETs serve as qubit reservoirs. The increased energy broadening of the SET reservoir becomes relevant, as soon as the readout scheme relies on the reservoir for the spin initialization, such as PSB readout of a freshly loaded electron. Hence, our results provide a strong incentive to separate the initial loading of an electron by adopting a readout scheme which does not rely on spin initialization via a reservoir. The initialization of an isolated PSB setting has for example been realized via projection and manipulation of the spin.\cite{Philips.2022}

\begin{acknowledgments}
 This work was funded by the German Research Foundation (DFG) within the project 289786932 (BO 3140/4-2 and SCHR 1404/2-2) and under Germany's Excellence Strategy - Cluster of Excellence Matter and Light for Quantum Computing" (ML4Q) EXC 2004/2 - 390534769. Part of this work was supported by the German Federal Ministry of Research, Technology and Space within the joint project QUASAR and has been carried out within the Joint Lab “Spin-Based Quantum Computing” established between IHP - Leibniz Institute for High Performance Microelectronics, RWTH Aachen University and Forschungszentrum Jülich.
\end{acknowledgments}

\section*{Conflict of Interest}
The authors declare no competing interest.

\section*{Data Availability Statement}
The data that support the findings of this study are available from the corresponding author upon reasonable request.

\bibliography{main.bib}

\end{document}